\documentclass[aps, prd,showpacs, superscriptaddress, groupedaddress]{revtex4-2} 
\usepackage{graphicx}  
\usepackage{subcaption}
\usepackage{float}
\usepackage{tabularx}
\usepackage{array}
\usepackage{amssymb}
\usepackage{mathtools}
\usepackage{dcolumn}
\usepackage{color}
\usepackage[pagebackref=false, colorlinks=true]{hyperref}
\hypersetup{
linkcolor=blue,     
citecolor=blue,     
urlcolor=blue}   
\usepackage{amsmath}
\usepackage{ulem}
\usepackage[utf8]{inputenc}

\begin{document}
\title{Astrophysical Black holes: An Explanation for the Galaxy Quenching}

\author{Jay Verma Trivedi}
\email{jay.verma2210@gmail.com}
\affiliation{International Centre for Space and Cosmology, School of Arts and Sciences, Ahmedabad University, Ahmedabad-380009 (Guj), India.}
\author{Pankaj S. Joshi}
\email{pankaj.joshi@ahduni.edu.in}
\affiliation{International Centre for Space and Cosmology, School of Arts and Sciences, Ahmedabad University, Ahmedabad-380009 (Guj), India.}
\author{Gopal-Krishna}
\affiliation{UM-DAE Centre for Excellence in Basic Sciences (CEBS), Vidyanagari, Mumbai - 400098, India.}
\email{gopaltani@gmail.com}
\author{Peter L. Biermann}
\affiliation{Max Planck Institute for Radio Astronomy, D-53121 Bonn, Germany.}
\affiliation{Department of Physics \& Astronomy, University of Alabama, Tuscaloosa, AL 35487, USA.}
\email{plbiermann@mpifr-bonn.mpg.de}

\date{\today}
\begin{abstract}
\textcolor{black}{In light of increasing observational evidence supporting the existence of ultra-compact objects, we adopt the term astrophysical black hole to refer to any object having a huge mass confined within a sufficiently small region of spacetim\textcolor{black}e. This terminology encompasses both the classical black hole solutions predicted by general relativity, as well as alternative compact objects that may not possess an event horizon.} We propose models of Astrophysical Black holes (ABHs) without event horizons (EHs), as a more viable explanation for the long-term quenching phenomenon in galaxies. At the same time, the short-term quenching is explained here in terms of an efficient feedback expected in the models of stellar-mass astrophysical black holes ({StMABHs}).
We have calculated the radiative flux from the disk in a general spherically symmetric metric background and used it to contrast the distinctive features of the BHs and ABHs scenarios.
\textcolor{black}{We demonstrate the relative ease of wind generation from the accretion disk surrounding an ABH without an event horizon, compared to a BH, and highlight the significant strength of these winds.}
{The nature of the feedbacks arising from accretion onto a BH and an ABH in the `quasar' and `radio' modes are compared}
and some possible observational signatures of the {StMABHs} are {pointed out}.\\

\textbf{keywords}: Galaxy Quenching, Astrophysical Black holes, Ultra-compact Objects, Accretion,  JWST.

\end{abstract}
\maketitle
\section{Introduction}
Wide-field galaxy surveys have shown that various properties of galaxies, such as optical color, star formation rate (SFR), and specific SFR (sSFR), are distinctly separated into two groups. In physical terms, these correspond to galaxies experiencing active star formation\cite{kronberg1, kronberg} and those without a significant ongoing star formation (``quenched galaxies'').
Remarkably, even in quenched galaxies, most baryons are not found locked within stars\cite{forman,fabian}. Instead, they are typically part of a hot gaseous halo surrounding the galaxy\cite{halo, fabian2, voigt, mcnamara}.
Thus, the quenching does not occur because the gas supply needed for star formation has run out, for there is plenty of gas in the halo.
A related issue arising from this, which has been widely discussed, is the `cooling problem': why doesn't the halo gas cool down and fall on to the host galaxy, causing an enhanced star formation there?\cite{fabian2, fabian3, fabian4}.
Empirically understanding the main observational factors linked to the transition from the star-forming to quenched state, has become a major focus in the study of galaxy formation and evolution through the cosmic history.

Over the last few decades, feedback has emerged as a promising resolution to the puzzle of low star-formation efficiency observed in many galaxies.
The feedback itself encompasses a broad spectrum of intricate processes, going beyond gravitational effects, and these collectively influence/govern the star formation \cite{bluck}. 
In contemporary numerical simulations, feedback is typically sub-divided into two broad categories: (i) feedback from stars and supernovae, as highlighted in the studies such as \cite{cole}, and (ii) feedback stemming from the accretion by the supermassive black hole (SMBH) in the active galactic {nuclei} (AGNs)\cite{somerville}.
In semi-analytic models, feedback from supernovae is typically invoked to effectively regulate star formation in low-mass galaxies (with stellar masses below $10^{10.5}$ $M_\odot$). Conversely, AGN feedback is leveraged to effectively suppress star formation in high-mass galaxies
\cite{bower,sijacki,vogelsberger,schaye,henriques}. Moreover, some theoretical models also incorporate environmental effects, such as ram pressure stripping of gas from the satellite galaxies in clusters and the ``strangulation" of gas supply, as considered in some other studies (e.g. \cite{vogelsberger2,henriques2}).

Supermassive black holes (SMBHs) situated at the center of galaxies go through periods of intense matter accretion\cite{llja}. During these phases, they radiate profusely and are identified as active galactic nuclei (AGN). AGN inject large amounts of energy into their host galaxies, approximately 10\% of the rest-mass energy of the accreted matter \cite{lamperti,shapiro, marconi}. This energy release can potentially impact the physical state of the interstellar medium (ISM) and, consequently, the star formation rate (SFR), a process referred to as AGN feedback\cite{fabian4, king, harrison}. In the simulations of galaxy formation, AGN feedback plays a vital role in the case of massive galaxies, in order to reproduce their varied observational features, such as the evolution of galaxy sizes\cite{choi}, the bimodality of galaxy colors\cite{trayford}, and the absence of very massive galaxies even inside the extremely massive galaxy haloes\cite{somerville2, behroozi}.
Both fast winds $( >1000 km/s)$ and collimated radio jets from the AGN accretion disk can serve as effective AGN feedback mechanisms, 
generating outflows that extend across the host galaxy over large distances\cite{king2, mukherjee, costa}{(these jets routinely go far beyond the outskirts of the most massive galaxies, with typical scale of 300 kpc size, with record many Mpc.)}. These outflows of thermal and relativistic plasma from the nuclear region can heat the interstellar medium (ISM) and expel the gas from the host galaxy\cite{ishibashi, zubovas}.

Recently, researchers have increasingly employed machine learning technique to analyze astronomical data and thus identify causal relationships\cite{brownson}. The strong correlation between black hole mass and bulge mass, indicates that galaxies with a larger bulge tend to harbour a more massive {central} black hole\cite{Magorrian:1997hw,Haring:2004hr}. This correlation might underlie the dependence of galaxy quenching on the bulge mass, as more massive black holes residing there can be expected to yield a greater amount of historical energy/momentum feedback into the galaxy {halo}, suppressing star formation more effectively.
The hypothesis gains further support from the superior predictive ability of the central velocity dispersion of the bulge, as compared to the estimated bulge mass, 
although this inference is limited to the spectroscopic data available. The tighter correlation of the central velocity dispersion with the black hole mass, compared to that of the bulge mass, could readily explain the higher central density, or central velocity dispersion found for the quenched galaxies\cite{brownson}.

In this paragraph, we outline the basic picture set out in the present study. The long-term quenching of galaxies is primarily attributed to a dense, compact object situated at the galactic nucleus\cite{report}. Conventional black holes possess event horizon, limiting matter accretion up to {a} boundary which is set at the ``Innermost stable Circular Orbit'' (ISCO) and the consequent {upper} bound on the generation of ultra-high-energy particles.
The cosmic censorship conjecture\cite{Penrose:1969pc, Hawking:1979ig} remains a foundational yet unproven and not even properly formulated, assumption in black hole physics. While it holds in simplified models such as spherically symmetric dust collapse, a wide range of studies on more general collapse scenarios—including those involving inhomogeneous matter fields, pressure, and rotation—have shown that naked singularities can emerge as legitimate solutions to Einstein’s equations \cite{Dwivedi:1992fh, Joshi:1993zg, Joshi:1994br, Joshi:2001xi, Hellaby:1985zz, Goncalves:2001pf, Giambo:2002xc, Harada:2001nj, Ori:1989ps, Gundlach:1999cu, Nolan:2002zd, Giacomazzo:2011cv, Ortiz:2011jw, Banerjee:2002sy, Barausse:2010ka, Mosani:2023vtr}. Despite multiple attempts, no rigorous mathematical proof exists for any version of cosmic censorship, and the conjecture’s vague formulation continues to hinder progress. In this light, the theoretical exclusion of naked singularities is no longer viable. Since the formation of such singularities is a generic outcome in many physically reasonable collapse models, the focus now necessarily shifts toward their observational signatures. \textcolor{black}{Recent studies have explored the possibility that some astronomical observations, including the first image of Sagittarius $A^*$, can be explained by naked singularity models, with the JMN-1 model \cite{Joshi:2011zm} recognized as a leading black hole mimicker for Sgr $A^*$ \cite{EventHorizonTelescope:2022xqj}. Other works have also considered naked singularities as valid candidates \cite{Ghasemi-Nodehi:2021ipd, Chakraborty:2017nfu}. Distinguishing naked singularities from black holes observationally is crucial and can be approached using various methods, including analyzing shadows-which may resemble black holes or appear as full-moon images depending on metric parameters \cite{Shaikh:2018lcc, Kaur:2021wgy}, perihelion precession, where naked singularities show both positive and negative precession unlike black holes \cite{Bambhaniya:2019pbr, Joshi2020, Bambhaniya:2022xbz}, and Lense-Thirring precession effects on orbiting gyroscopes or pulsar beams, which differ near naked singularities compared to black holes \cite{Bambhaniya:2021jum, Kocherlakota:2017hkn}. Additional techniques include measurements of relativistic time delays of pulsar signals \cite{Kalsariya:2024qyp} and tidal forces \cite{Joshi:2024djy}, all of which offer promising routes to verify the existence of naked singularities.}

In this work, we consider the astrophysical implications of naked singularities—viewed as ultra-compact, massive objects without event horizons—and explore their potential impact on galaxy evolution, particularly in comparison to traditional black holes. To address this, we propose here models of ``astrophysical black holes'' (ABHs) that are devoid of {an} event horizon (EH), hence characterized by a huge amount of matter confined within an extremely compact region, a situation which permits the disk of accreting matter to extend all the way to the centre and yield extremely high energy densities and temperatures. This access to the ultra-strong gravity regime facilitates generation of extremely energetic particles, thus contributing to the long-term quenching of star formation in the host galaxy and its halo region. Additionally, we introduce models of ``stellar-mass astrophysical black holes'' ({StMABHs}) which can arise from gravitational collapse of massive stars once they run out of their internal nuclear fuel. These {StMABHs}, prior to {getting} enveloped by the event horizon, create high-density regions capable of emitting very high-energy particles over short time scales. Such a situation may lead to short-term episodes of (localised) {intra-galactic} quenching. Potential observational signatures associated with the {StMABHs} are also pointed out here. Section II provides a brief overview of the recent observations related to this phenomenon, made with the James Webb Space Telescope (JWST). Section III touches upon the role SMBHs play in the galaxy quenching process. Section IV presents some salient features of our proposed models involving astrophysical black holes (ABH), in the context of long- and short-term quenching. Our conclusions are summarised in Section V.

\section{Galaxy Quenching: Recent clues from the JWST observations}
Among the most critical unsolved problem in galaxy evolution is the issue of onset of star formation in primordial haloes, leading to galaxies in the early universe and the question of subsequent cessation of the star-forming activity in such young galaxies. The release of initial data from the James Webb Space Telescope (JWST) has led to a remarkable advance in addressing the former of these questions. Throughout the evolution of the Universe, from the cosmic noon to the present era, a significant, steady decline has been observed in the conventionally estimated rates of star formation in galaxies. As one explores higher redshifts, beyond the cosmic noon, galaxies experiencing active star formation emerge as predominant entities.

Discovery of 10 massive quiescent galaxies at redshifts greater than 3 was reported in the initial data released under the JWST Cosmic Evolution Early Release Science program\cite{carnall}. Three of these galaxies lying in the redshift range $z = 4$ to $5$, provide the most compelling evidence hitherto for the existence of quiescent galaxies prior to $z = 4$ (this is in reference to the results of \cite{carnall} where they found that few of the galaxies are quenched earlier than $z=4$). These remarkable galaxies have estimated stellar masses (i.e. $log_{10}(M_*/M_\odot)$, where $M_*$ represents the total stellar mass of the galaxy) in the range from $10.1$ to $11.1$, with their major stellar mass accumulation having occurred around $z \sim 10$. Furthermore, the star formation histories of two of these galaxies suggest that they had already attained $log_{10}(M_*/M_\odot)>10$ by $z \sim 8$. Previously, quenched galaxies had only been found up to $z \sim 5$, exhibiting traits of both a significant mass $(M_* > 10^{10} M_\odot)$ and an advanced age\cite{Glazebrook, Valentino, Nanayakkara}. However, the problem has now been accentuated by the recent discovery of
a mini quenched galaxy at $z = 7.3$, when the universe was barely 700 million years old. This galaxy underwent a brief starburst phase followed by a swift quenching phase \cite{looser}. With an estimated stellar mass of just $4-6 \times 10^8 M_\odot$, the galaxy falls in a regime where galaxy formation is susceptible to multiple feedback processes, potentially resulting in only a temporary quenching outcome, because a sustained quenching due to feedback becomes realizable only when a fairly massive black hole has formed, an imperative which necessitates availability of a densely concentrated central core \cite{report}.


\section{The role of Supermassive Black holes (AGN)}
The phenomenon of galaxy quenching can be linked to multiple causes and processes. Galaxies may cease star formation either due to the {expulsion} of their gas (the raw material), which would render it incapable of sustaining star formation, or because the gas cannot be transported from the surrounding halo into the main body of the forming galaxy.
The gas {expulsion} could occur through outflows resulting from AGN activity, potentially {depleting} the molecular gas reservoir of the host galaxy. However, simulations and observations indicate that AGNs tend to inhabit galaxies with abundant molecular gas content\cite{simulation, obs}, complicating the overall perspective. Thus, while AGNs can generate outflows potentially capable of inducing quiescence
 {in the host galaxy}, they are often seen to exist in galaxies rich in molecular gas, implying that the basic raw material for star formation is in fact available\cite{report}. Another {signature of AGN feedback is}
AGN dust heating, which contributes to far-infrared emissions. Investigating AGN demographics across the cosmic history could provide valuable insight into the galaxy quenching processes. A yet another mode of AGN feedback involves  the generation of massive galaxy-scale winds that would heat the gas within the surrounding halo, primarily along the minor axis of the central galaxy\cite{minor}. {Furthermore}, these winds can expel gas from the inner regions, diminishing the availability of cooler gas for the star formation process\cite{report}.
In view of the prevailing lack of consensus about the mechanism(s) for the observed rapid quenching of star formation in galaxies in the early universe, we explore in the next section some plausible alternative scenarios, with the aim to explain both short-term (temporary) and long-term quenching of star-formation activity in galaxies.

\section{Astrophysical Black holes: A potentially more viable explanation}
When referring to astrophysical black holes(ABHs), we mean an enormous quantity of matter confined within an extremely compact region (see below). It is widely recognized that massive galaxies typically harbour a compact object of very high density and mass at the center. Conventional mathematical description of such supermassive black holes posits the presence of an event horizon, defined as the boundary marking the causal past of future null infinity in a given general relativistic spacetime geometry. Consequently, when matter accretes towards such black holes, the accretion disk extends only up to the Innermost Stable Circular Orbit (ISCO), which is well outside the event horizon, at $3R_s$, where $R_s$ is the Schwarzschild radius, the location of the event horizon. The region within the event horizon, containing trapped surfaces, prohibits any expulsion of matter, {by} directing all material within it towards the singularity as its inevitable endpoint. As a result, the accretion {directed towards} such black holes cannot continue beyond a certain limit, with a natural cut-off imposed by the ISCO. Consequently, the winds or high-energy particles emanating from such an accretion disk may not attain very high power.


\subsection{A plausible scenario for long-term quenching}
Long-term quenching can be attained through a {sustained} vigorous emission of high-energy particles emanating from the central entity of the galaxy. Suppose the central entity is an astrophysical black hole, an ultra-compact object lacking an event horizon. In that case, the accretion disk will extend toward the centre,  as opposed to its terminating at the ISCO (see above). As we shall argue, this scenario has the potential to offer a more plausible and realistic explanation for the long-term quenching, as it permits extremely high-energy {particle} emissions over a long period of time. In 2011 and 2014, two such models were introduced, each suggesting the formation of an ultra-dense region through the {dynamical} collapse of certain distinct types of matter, such as a perfect fluid and others. These Joshi-Malafarina-Narayan(JMN) models posit \textcolor{black}{naked singularities, which are the} astrophysical black holes (ABHs) that lack an event horizon, thus allowing the accretion disk to extend essentially right up to the center\cite{Joshi:2011zm,Joshi:2013dva}. 

The paper JMN1\cite{Joshi:2011zm} describes the gravitational collapse of an anisotropic fluid, characterized by {zero radial pressure} and non-zero tangential pressures. The matter undergoes collapse, eventually reaching an equilibrium configuration asymptotically, where the central region harbors a singularity devoid of a horizon. This equilibrium state is achieved over an infinite comoving time span. However, when the matter's velocity during the collapse approaches a minimum, the central region gets transformed into an ultra-dense compact object, an ``astrophysical black hole'' without any trapped surfaces forming, with the central density increasing progressively with ongoing collapse. During this phase, the metric of the collapsing fluid in co-moving coordinates $(t,r,\theta,\phi)$ is, 
\begin{equation}\label{jmn1}
    ds^2 = - e^{4\int_0^\omega \omega \frac{R'}{R} dr} dt^2 + \frac{R'^2}{b(r)e^{4\int_v^1 \frac{\omega}{v} dv}} dr^2 + R^2 d\Omega^2.
\end{equation}
where $\omega$ is the equation of state of the matter, defined as $\omega=p_\theta/\rho$, physical radius $R$ is defined as $R(r,t)=r v(r,t)$, and $b(r)$ is a free function. 
A comprehensive examination of stable circular orbits and their comparison with the Schwarzschild solution is presented in \cite{Joshi:2011zm}, revealing a discernible contrast between the two frameworks. Notably, the discussion sheds light on the phenomenon of stable circular orbits, or the accretion disk extending towards the central region.

\begin{figure}[h!]
    \centering
    \includegraphics[width=8cm, height=6cm]{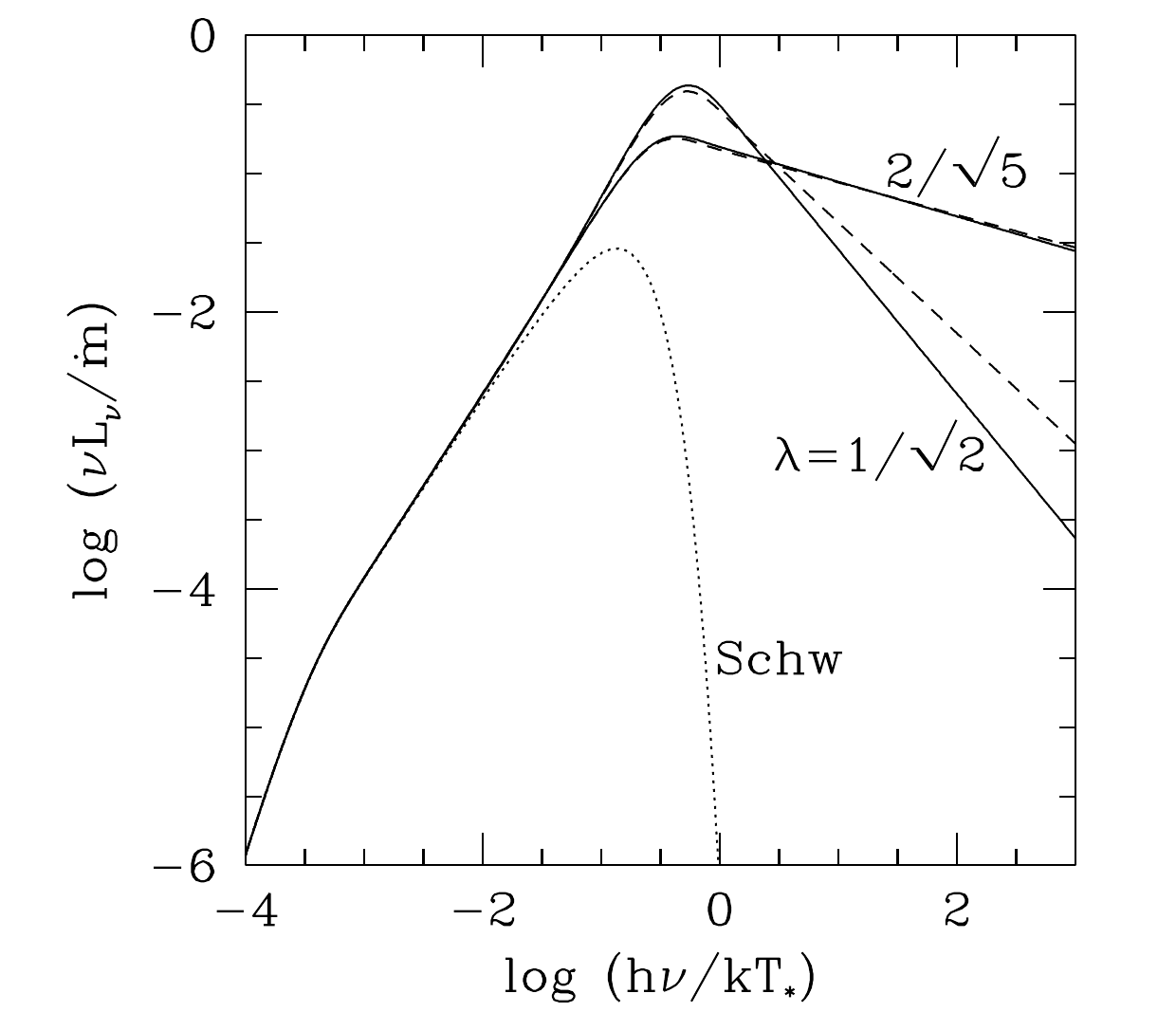}
    \caption{Spectral luminosity distribution of radiation from the accretion disc. The dotted line corresponds to a disc around a Schwarzschild black hole, the solid lines correspond to discs around the two perfect fluid ABH without EH models, as discussed in \cite{Joshi:2013dva}. The dashed lines are for the discs around two tangential pressure ABH without EH, as modelled in \cite{Joshi:2011zm}. Here, $\lambda$ is the model-specific parameter that relates to the masses of the perfect-fluid ABH, given in \cite{Joshi:2013dva}. 
    {$T_*$ is the characteristic temperature of the local patch of the disc radiating as a blackbody.}}
    \label{joshi2014}
\end{figure}

A slowly evolving, collapsing cloud of perfect fluid can asymptotically settle to a static spherically symmetric equilibrium configuration, with a singularity at the center, as discussed in JMN2\cite{Joshi:2013dva}. The metric of this collapsing fluid in the co-moving coordinates has the form,
\begin{equation}\label{jmn2}
    ds^2 = - e^{-2\int_0^r\frac{p'}{\epsilon+p} dr} dt^2 + \frac{R'^2}{b(r)e^{2\int_r^R\frac{v'}{R'} dR}} dr^2 + R^2 d\Omega^2.
\end{equation}
The fundamental observational characteristics of accretion disks orbiting within the simplified models of equilibrium configurations derived from the above family of spacetimes are investigated in \cite{Joshi:2013dva}. That analysis assumes the accretion disk to be gaseous, where the loss of angular momentum due to viscosity results in a gradual inward motion along a series of nearly circular orbits.
The calculation of the luminosity, denoted as $L_\infty$ (energy per unit time), which reaches an observer located at infinity, is conducted to facilitate the estimation of the spectral luminosity distribution of the radiating accretion disc. Fig. (\ref{joshi2014}) illustrates the disparity/difference between the emissions from the disc surrounding conventional black holes with event horizons and those from the astrophysical black holes mentioned above.
A pronounced cutoff in the emissions due to the presence of an event horizon is evident for the case of the Schwarzschild black hole. 
Conversely, the ABH models exhibit high-energy emission tails.

\subsection{Temporary Quenching}\label{section ltb}
As discussed in the previous section, an astrophysical black hole without an event horizon located at the center of a galaxy, may contribute to the long-term quenching of its star formation. On the other hand, short-term quenching (along with the starbursts) could result from individual stellar-mass astrophysical black holes formed through the gravitational collapse of sufficiently massive stars.
In the 1930s, Oppenheimer and Snyder, and independently B. Datt, described the gravitational collapse of a spherically symmetric homogeneous dusty star, characterized by zero pressure, and elucidated the process through which the Schwarzschild black hole is formed\cite{Oppenheimer:1939ue, datt}.
In the 1990s, a more realistic scenario was conceptualized, which involved the collapse of a star, described by Tolman-Bondi spacetimes\cite{Dwivedi:1992fh, Joshi:1993zg}. This approach posed two intriguing possibilities: one resulting in the conventional black hole and the other entailing a delay in the formation of both the event horizon and the apparent horizon. In the latter case, the horizons form at the instant central singularity forms, momentarily rendering the singularity visible in comoving coordinates. It was demonstrated that an infinite family of null geodesics can emanate from the immediate proximity of the singularity and extend to reach observers located at infinity\cite{Dwivedi:1992fh, Joshi:1993zg}. Quite possibly, the radiation emerging along these geodesics may be sufficiently energetic to induce temporary quenching within the galaxy. A quantitative estimate of this energy would require the knowledge of the theory of quantum gravity, or it may require a detailed phenomenological analysis (see also \cite{Goswami:2005fu}).

\subsection{The Role of Radiative and Mechanical Efficiency}
Black holes with super-Eddington accretion 
would operate in a radiatively efficient, `quasar' mode. In contrast, BHs with {highly} sub-Eddington accretion operate in a radiatively inefficient, `radio' mode. Despite the lower radiative efficiency, this `radio' mode generates substantial mechanical luminosity, driving winds and/or collimated jets\cite{Sijacki:2007rw}. The magnitude of the Eddington ratio determines the transition between the two accretion (i.e., feedback) modes. We recall the simplified treatment by Silk and Rees\cite{Silk:1997xw}, where they estimated the effect of the accretion disk winds on the galaxy. The galaxy was assumed to be an isothermal sphere having a gas fraction, $f_{gas}$ and a density $\rho=v^2/2\pi G r^2$, where $v$ is the velocity dispersion. \textcolor{black}{As discussed in \cite{Silk:1997xw}}, the wind can propel the interstellar gas out of the galaxy with a constant velocity,
\begin{equation}
     v_s = \left(\frac{f_w L_{Edd} 8\pi^2G}{f_{gas}v^2}\right)^\frac{1}{3}.
\end{equation}
Here, the wind luminosity is the fraction $f_w$ of the Eddington luminosity. The fraction $f_w$ is dependent on the radiation efficiency $\epsilon$ as\textcolor{black}{\cite{Silk:1997xw}},
\begin{equation}
    f_w \equiv \frac{\dot{M}_{out}v_w^2}{L_E} = \frac{1}{\epsilon}\left(\frac{v_w}{c}\right)^2,
\end{equation}
where $\epsilon=L_E/\dot{M}_{inf}c^2$. For $v_s>v$, 
\begin{equation}\label{eqn 5}
    M_{BH} > 8\times10^8 \gamma \left(\frac{v}{500~km s^{-1}}\right)^5 M_{\odot},
\end{equation}
where $\gamma \equiv f_{gas}/32\pi^3 f_w$. In theory then, a black hole could expel all the material out of its host {galaxy}, once its {mass} surpasses $\sim 10^7 M_{\odot}$(for $\gamma \sim 1$ \& $f_w \sim 0.01$){(however, these arguments hold for early galaxies)}. 

When analysing the accretion disk around a BH, assuming that the gas in the disk loses angular momentum due to viscosity and gradually moves inward along a series of nearly circular orbits, one can determine the radiative efficiency. By considering only the characteristics of these circular geodesics, without requiring detailed knowledge of the viscous stress properties, it is possible to determine the radiative flux emitted at each radius within the disk\cite{Page:1974he, Novikov:1973kta}. 
First, we discuss the general derivation of the radiation efficiency for the general metric
\begin{equation}
    ds^2 = -g_{tt}(t,r) dt^2 + g_{rr}(t,r) dr^2 + g_{\theta\theta}(t,r) d\theta^2 + g_{\phi\phi}(t,r) d\phi^2.
\end{equation}
The energy per unit mass and the angular momentum per unit mass can be calculated using the two {Killing} vectors associated with time-translational symmetry($\zeta \equiv \{1,0,0,0\}$) and rotational symmetry($\eta \equiv \{0,0,0,1\}$). The energy $E=g_{\alpha\beta}\zeta^\alpha u^\beta$ and the angular momentum $L=g_{\alpha\beta}\eta^\alpha u^\beta$ can be expressed in terms of the metric functions for the timelike geodesics. The four-velocity of the timelike geodesics follow $u^\alpha u_\alpha=-1$, from which, we can get,
\begin{equation}\label{timelike}
    E^2 = V_{eff} + g_{tt}g_{rr} \left(\frac{dr}{d\tau}\right)^2,
\end{equation}
where, $V_{eff}$ is the effective potential having the form,
\begin{equation}\label{8}
    V_{eff} = g_{tt}\left(1 + \frac{L^2}{g_{\phi\phi}}\right).
\end{equation}
Requiring $\partial V_{eff}/\partial r = 0$ for the circular orbits, we get,
\begin{equation}\label{9}
    L^2 = \frac{g_{\phi\phi}^2g'_{tt}}{g_{tt}g'_{\phi\phi}-g'_{tt}g_{\phi\phi}}.
\end{equation}
Where {superscipt} $'$ represents the partial derivative with respect to $r$. Setting $dr/d\tau = 0$ in Eqn. (\ref{timelike}), for a circular orbit, we get
\begin{equation}\label{10}
    E^2 = \frac{g_{tt}^2g'_{\phi\phi}}{g_{tt}g'_{\phi\phi}-g'_{tt}g_{\phi\phi}}.
\end{equation}
The angular velocity of these circular orbits can be expressed as,
\begin{equation}\label{11}
    \omega^2 = \left(\frac{d\phi}{dt}\right)^2 = \frac{g'_{tt}}{g'_{\phi\phi}}.
\end{equation}
Using these $L$, $E$, \& $\omega$, we can calculate the radiative flux
emitted by the disc (which is the energy per unit area per unit time) in the local frame of the accreting fluid as\cite{Page:1974he, Novikov:1973kta},
\begin{equation}\label{12}
    \mathcal{F}(r) = -\frac{\Dot{m}}{4 \pi \sqrt{-g}} \frac{\omega'}{(E-\omega L)^2} \int^r_{r_{in}} (E-\omega L)L' dr.
\end{equation}
Here, the {superscipt} $\Dot{}$ represents the partial derivative with respect to $t$ and $g$ represents the determinant of the metric, $g=-g_{tt}g_{rr}g_{\theta\theta}g_{\phi\phi}$. $r_{in}$ is the innermost stable circular orbit, which can be found by requiring,
\begin{equation}\label{13}
    \frac{\partial^2 V_{eff}}{\partial r^2}\bigg|_{r=r_{in}} \geq 0.
\end{equation}
A more practical measure is the luminosity, i.e. the energy that reaching an observer at infinity $\mathcal{L}_\infty$, per unit time. By integrating $d\mathcal{L}_\infty/d ln r = 4 \pi r \sqrt{-g} E \mathcal{F}$ with $ln r$, we can calculate the net luminosity $\mathcal{L}_\infty$ observed at infinity. 
As the disc is truncated at the ISCO in the BH case, the efficiency is only $\sim 6\%$ as $\epsilon = 1-E_{ISCO} = \mathcal{L}_\infty/\dot{m}c^2$. In the case of ABHs (see, JMN1 \& JMN2), the accretion disk can extend up to the central high-density entity. However, the singularity is not yet formed, so the accretion disk {extends} up to the boundary of the central ultradense object, and the energy of the innermost circular orbit will
{then} always be lesser than the $E_{ISCO}$. This essentially means that, as the collapse proceeds, the radiative efficiency of the ABH rises, \textcolor{black}{which can be verified by calculating $\mathcal{L}_\infty$ for the dynamic metrics given in (\ref{jmn1}) \& (\ref{jmn2}) using Eqns.(\ref{timelike})$-$(\ref{13})}. When the singularity has formed in the asymptotic time, the stable circular orbits can extend right up to the singularity. The radiative efficiency will then approach $100 \%$. The $\gamma$ of the Eqn. (\ref{eqn 5}) will be $\sim 17$ in that case, and we require $17$ times more mass than the typical BH, in order to expel all the material from the host galaxy. 
{This estimate is meant for the feedback due to mechanically generated winds from the accretion disk. The calculation assumes that the mechanical efficiency is reversibly proportional to the radiative efficiency. The radiative efficiency of ABH being very high, it would have a lower efficiency for mechanical energy injection and hence required to be more massive. However, when the winds are accelerated by the disc radiation, 
the ABH will be able to generate high-energy winds with greater efficiency.}
The next subsection discusses this in detail with the example of the hydrodynamical simulation for a naked singularity spacetime. In the BH case, the radiation-dominated (quasar) mode sets in only when the accretion rate becomes very high. {In contrast, an} ABH can produce very efficient radiation-driven winds even when the accretion rate is very low. 

The distinction between the BH and ABH cases becomes evident when their radiative and mechanical efficiencies in the different accretion modes are 
{compared}. In the quasar mode, where the mass accretion rate is very high, the central ultra-dense region of the ABH is forming, so the disc will be more efficient in generating mechanical winds (i.e., jets, etc.), as compared to producing radiation-driven winds composed of the in-falling matter, because the radiative efficiency is increasing. 
Once the central ultra-dense region has formed, and the mass accretion rate is very low (i.e., radio mode), the ABH would still be more efficient in producing radiation-driven winds in comparison to BH, and this will be reflected in stronger radiation-driven winds and jets even in the radio mode. 
\subsection{Wind Generation}
\textcolor{black}{We can get the idea of the wind generation due to the effects of the radiation pressure through the mechanism discussed in \cite{winds}. The key assumptions of the mechanism are as follows: Particles are ejected from the surface of a geometrically thin disk around a central object of mass $M$. The gravitational force of the central object, the centrifugal force, and the radiative forces from the disk act on the particles. The gravitational effects of the disk gas are neglected. The radiation force, the radiation drag from the central object, and the radiation drag due to the disk radiation field and the very high velocity motion of the particles are also considered zero. The near disk approximation is used for simplicity, in which the particles are assumed to suffer from the radiation fields of the disk just below them. The motions of the particles are calculated using the Newtonian formalism. In the cylindrical coordinates $(r,\phi,z)$, where $z$ axis is along the axis of the rotation of the disk, the equations of motion for the particles are\cite{winds},
\begin{equation}\label{vr}
    \frac{dv_r}{dt} = \frac{v^2_\phi}{r}-\frac{GMr}{R^3},
\end{equation}
\begin{equation}\label{vp}
    \frac{1}{r}\frac{d}{dt}(rv_\phi)=0,
\end{equation}
\begin{equation}\label{vz}
    \frac{dv_z}{dt}=-\frac{GMz}{R^3}+\frac{\sigma_T}{mc}F_d.
\end{equation}
Here, $R=\sqrt{r^2+z^2}$ and $F_d$ represents the radiation flux from the disk and can be expressed as,
\begin{equation}
    F_d=\sigma T^4_d=\frac{3GM\dot{M}}{8\pi r^3}\left(1-\sqrt{\frac{r_{ISCO}}{r}}\right),
\end{equation}
where, $\dot{M}$ represents the constant accretion rate, $r_{ISCO}$ is the inner most stable circular orbit and $T_d$ is the effective temperature of the disk. 
The Eqn.(\ref{vr}) represents the equation of motion in the radial direction, with the first term being the centrifugal force and the second being the gravitational force due to the central object, on the right-hand side. The Eqn.(\ref{vp}) is the angular momentum conservation equation. The Eqn.({\ref{vz}}) is the equation of motion in the vertical direction, with the first term being the gravitational force due to the central object and the second being the radiation force due to the disk radiation, on the right-hand side. We can define the normalized disk luminosity as},
\begin{equation}\label{gama}
    \Gamma_d=\chi_e\frac{L_d}{L_E},
\end{equation}
with ionization rate $\chi_e=1$, Eddington luminosity $L_E=4\pi G   cMm_p/\sigma_T$ (here $m_p$ is the proton mass), and the disk luminosity\textcolor{black}{\cite{winds}},
\begin{equation}\label{ld}
    L_d=\frac{GM\dot{M}}{2r_{ISCO}}.
\end{equation}
In terms of $\Gamma_d$, Eqn.(\ref{vz}) can be rewritten as,
\begin{equation}
    \frac{dv_z}{dt}=-\frac{GMz}{R^3}+\frac{3\Gamma_dGMr_{ISCO}}{r^3}\left(1- \sqrt{\frac{r_{ISCO}}{r}}\right).
\end{equation}
To understand wind generation, we need to examine the escape condition for the particles. This can be achieved using the net specific energy of the particles ejected from the disk. For the particles ejected from the disk with the initial conditions: $v_r=0$, $v_\phi=v_{Keplerian}=\sqrt{GM/r}$, $v_z=0$, and $r=r_0$ at $z=0$ this specific energy will be of the order of
\begin{equation}\label{ene}
    E=\frac{v^2_\phi}{2}-\frac{GM}{r_0}+\frac{3\Gamma_dGMr_{ISCO}}{r_0^3}\left(1- \sqrt{\frac{r_{ISCO}}{r_0}}\right)z.
\end{equation}
The right-hand side consists of the rotational energy in the first term, the gravitational potential energy in the second, and the work done by the radiation force of the disk. If we measure $r_0$ in terms of $2GM/c^2$ and velocity in terms of $c$, the Eqn.(\ref{ene}) will take the form,
\begin{equation}
    E=-\frac{1}{2r_0}+\frac{3}{2}\Gamma_d\frac{r_{ISCO}}{r_0^3}\left(1- \sqrt{\frac{r_{ISCO}}{r_0}}\right)z.
\end{equation}
The gravitational acceleration is largest when $z \sim r_0$, hence demanding this energy to be positive at $z \sim r_0$ gives the escape condition as,
\begin{equation}
    \Gamma_{d~escape}=\frac{1}{6}\frac{r_0}{r_{ISCO}}\frac{1}{\left(1- \sqrt{\frac{r_{ISCO}}{r_0}}\right)}.
\end{equation}
\begin{figure}[!h]
  \begin{subfigure}[!h]{0.45\textwidth}
    \includegraphics[width=\textwidth]{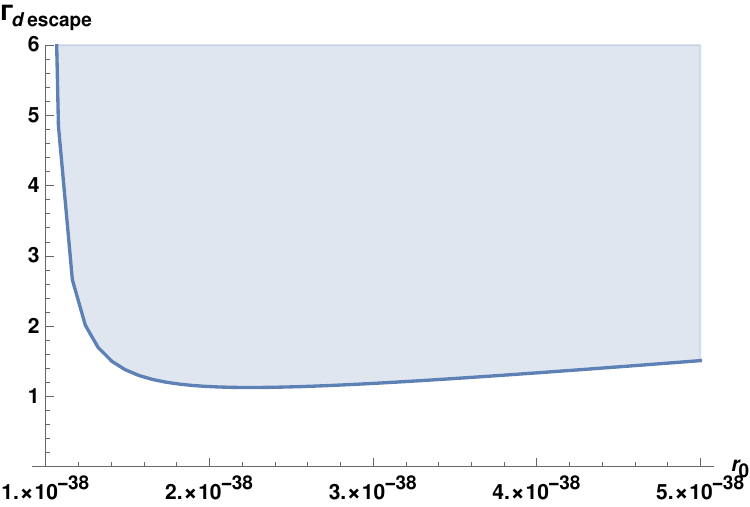}
    \caption{}
    \label{fig:f1}
  \end{subfigure}
  \hspace{0.5cm}
  \begin{subfigure}[!h]{0.45\textwidth}
    \includegraphics[width=\textwidth]{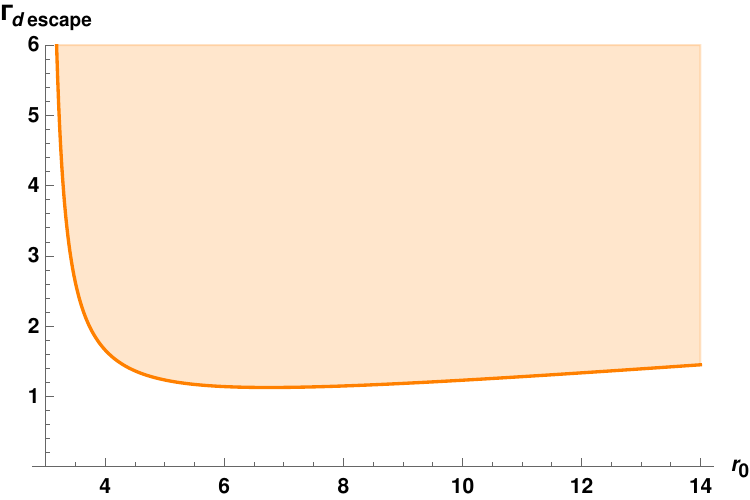}
    \caption{}
    \label{fig:f2}
  \end{subfigure}
  \caption{The shaded regions show the normalized luminosity $\Gamma_d$ for which the particles will escape from their respective positions $r_0$. (a)Escape condition for ABH without EH case (b)Escape condition for BH case.}
\end{figure}
In the units of $2GM/c^2$, the ISCO for the Schwarzschild black hole is located at $r_{ISCO}=3$. However, the ISCO for the ABH without EH models discussed in the previous section would be very close to the central ultra-dense object. If we assume the ultra compact object to be of the size of the order of Planck length, then ISCO for a few solar mass ABH without EH will be at $r_{ISCO}=10^{-38}$. This value is inversely proportional to the mass of the ultra-compact object. The behavior of the escape condition for BH and ABH without EH shows a similar trend of the required disk luminosity diverging as the position of the particle reaches the ISCO, as shown in the Figs.\ref{fig:f1} \& \ref{fig:f2}. However, as the disk extends to the central ultra compact object in ABH without EH, the divergence of the normalized luminosity occurs in the ultra strong gravity region compared to the Schwarzschild black hole where this divergence occur at the ISCO, which is very far from the EH. It is evident from Eqns. (\ref{gama}) \& (\ref{ld}), when the disk luminosity is low due to the low accretion rate, the wind generation from the BH is not possible. However, even with the very low accretion rate, the ABH without EH will be able to generate the winds due to their extremely small value of the ISCO and hence the large value of $\Gamma_d$. The Eqn.(\ref{gama}) shows that the $\Gamma_{d(ABH ~without~EH)}\sim3\times10^{38}\Gamma_{b(BH)}$ for the same rate of accretion. Hence, the winds generated from the accretion disk around the ABH without EH will be much stronger compared to those generated from the disk around a BH. These calculations neglect general relativistic effects; however, they provide a rough estimate of the relative ease of wind generation from ABHs without an event horizon compared to BHs, and consequently, their role in quenching star formation.
\textcolor{black}{Particle dynamics in the ultra-strong gravitational regime near ABH cannot be accurately modeled using the Newtonian framework applied in wind generation analyses. Although such approaches may provide approximate estimates, a precise treatment necessitates general relativistic numerical simulations, which we intend to pursue in future work. Previous hydrodynamical and general relativistic magnetohydrodynamic (GRMHD) studies of select naked singularity models have been conducted by other researchers \cite{Kluzniak:2024cxm, Uniyal:2025hik}, and these works are reviewed in the following section.}

\subsection{Outflows from Naked Singularities}
Further insight about the generation of outflows or jets from the naked singularities can be adduced via hydrodynamical simulations. Recently, researchers have performed hydrodynamical simulations of accretion onto Reissner-Nordström(RN) naked singularities and compared them with RN black holes\cite{Kluzniak:2024cxm}. The results were completely dependent on the existence of the event horizon, which {governs} the accretion pattern. That was ensured by the consideration of an electrically neutral fluid, which will be independent of the amplitudes of charges of RN spacetimes under study. The typical behavior of the black hole event horizon is observed. All the matter being accreted will be absorbed by the event horizon; hence, the outflows from the black holes can only be generated with the presence of a magnetic field, radiation, or rotation. However, RN naked singularities were observed to be favorable for the formation of outflows and jets. It was observed from the simulations that the fluid can have mildly relativistic velocities having turned around the RN naked singularities. This initial simulations of naked singularities clearly show the distinctive behavior of existence and non-existence of an event horizon. The mere fact that the accretion disk extends to the very central region of astrophysical black holes makes them favorable for generating outflows or jets, depending on the efficiency discussed in the previous subsection.
\textcolor{black}{However, the dynamics of real galaxies are governed by rotation, magnetic fields, and complex gas physics. Crucial astrophysical phenomena, such as jet launching, are fundamentally linked to both rotation and magnetic fields. Recent studies have employed general relativistic magnetohydrodynamic (GRMHD) simulations to investigate the accretion dynamics around JMN naked singularities \cite{Uniyal:2025hik}. These studies reveal that such naked singularities possess a centrifugal barrier that obstructs the direct accretion of rotating matter. Despite a reduction in magnetization within the funnel region, these singularities are capable of generating jet powers comparable to those observed in black holes. Moreover, strong outflows or jets emerge, with driving mechanisms that may vary with the specific spacetime geometry and spatial location—being dominated at times by gas pressure and at others by magnetic pressure. Importantly, the energy powering these outflows or jets is derived from the gravitational potential energy of the matter accumulated in the deep potential well. For particular parameter regimes, the JMN singularity can even produce jets exceeding the strength of those observed from black holes.}

\subsection{Accelerating Winds}
Recently, a signiﬁcant increase in the outflow velocity with time was detected in the spectrum of the quasar SBS $1408+544$ ($z = 2.337$)\cite{wind}. Using $\sim130$ spectra acquired over 8 years, as part of the spectroscopic monitoring during the SDSS, these authors have determined that the C{\small IV} broad absorption line (BAL) is not only varying in strength but also systematically shifting towards higher velocities. The net velocity shift is $683$ $kms^{-1}$, corresponding to a mean acceleration of $1.04$ $cm s^{-2}$ (however, the acceleration varied throughout the monitoring period). They consider a range of possible explanations for the observed velocity shift and argue that it {is} most likely explained either as a geometric effect, or as a real acceleration of the outﬂow, due to radiation pressure. In the second scenario, the observed acceleration of the BLR could be due to an increased speed of the outflow driven by a varying incident (ionizing) radiation from the disc\cite{Murray:1995zz}. Thus, they \cite{wind} estimate that a certain combination of BH mass and disk wind launch radius can reproduce the observed BLR acceleration. However, in this model, to have the acceleration increasing with time, either the wind originates predominantly at radii $r < 1.65 r_L$, since this model allows acceleration to increase only at such small radii, or the wind's terminal velocity is changing with time\cite{wind}. These constraints arise mainly due to the comparatively low efficiency of BHs in generating radiation from in-falling matter. In contrast, an ABH can easily explain this observation because its radiative efficiency increases with time and asymptotically approaches $100\%$, as discussed in the previous section. The increasing radiative efficiency is expected to accelerate the winds, exerting an increasing amount of radiation pressure. A detailed modelling of this construct, supported by numerical simulation of ABHs, is needed. 

\subsection{Some Observational Signatures}
Maximally rotating black holes can emit energy, fueled by their rotation, which {can be} several powers of ten above the Eddington rate {for stellar-mass black holes}, the typically observed power being around $10^{43.5}$ $erg/s$, {apparently} independent of BH mass \cite{EventHorizonTelescope:2019pgp}. 
This suggests that many young massive stars turn into maximally rotating BH\cite{Chieffi13, Limongi18, Limongi20} for a short time (about $10^4$ yr for a 10 solar mass BH, $M_{BH, 1}$, with the time scales rising linearly with BH mass) period.
For stellar mass BHs, as this energy output exceeds the Eddington luminosities by many powers of ten, their impact on the environment could be very substantial, assuming a sufficiently high star formation rate.
This observed power matches Eddington {luminosity} only for the smallest SMBHs. Such estimates have been published for over 25 years (e.g., \cite{Allen:1998vt,Punsly:2011pi}). In such a scenario, the star forming process itself does the required quenching, via freshly created BHs, and in case the initial rotation is higher than the limit allowed for freshly formed stellar-mass BHs, the disruption caused due to the feedback would be even {more pronounced}. This would indicate that freshly formed stellar mass BHs are not BHs with an event horizon, but they are ABHs (which would take about $10^4$ $M_{BH, 1}$ years to be covered within an event horizon). \textcolor{black}{Here we are referring to the ABH discussed in sec.\ref{section ltb}.} 
\textcolor{black}{In \cite{Allen:2024ibk}, the final structural configuration of a massive star immediately prior to its collapse and the subsequent formation of a rotating black hole is presented. The portion of the stellar core destined to form the black hole possesses approximately $60\%$ more angular momentum than the theoretical upper limit. We propose that, during the collapse, a transient naked singularity phase may occur, allowing the system to shed the excess angular momentum.} \textcolor{black}{To further assess this possibility, one would ideally investigate gravitational collapse scenarios through detailed numerical simulations and examine the quantum gravity effects expected in the vicinity of naked singularities. Such studies could help clarify the nature of radiation or matter that may escape from these objects and determine whether a connection with the observed galactic winds can be established.
Previous \cite{Goswami:2005fu} work has explored quantum gravity effects near naked singularities, particularly within the framework of loop quantum gravity. These studies indicate that quantum corrections introduced at the final stages of gravitational collapse can generate a strong outward energy flux. However, these analyses have so far been limited to specific collapse models, such as those involving scalar fields, and have not yet been extended to more general astrophysically relevant matter configurations.
At present, the absence of a complete and experimentally validated theory of quantum gravity prevents a fully rigorous investigation along these lines. Nevertheless, the scenario considered here may still be examined within an astrophysical context. In particular, one may search for possible quantum gravity signatures in the properties of these winds, which could eventually contribute to a phenomenological understanding of quantum gravity effects.}

Referring to the observations, the magnetic fields observed in terms of $(B \times r)$ $=$ $10^{16.0 \pm 0.12}$ $(G \times cm)$ in a wind that resembles the Parker wind\cite{parker} and carries powerful electric currents\cite{gopal} could actually be a (short-lived) signal for the putative transition from a compact object without a horizon to a BH with a horizon. \textcolor{black}{However, this remains a preliminary proposal; further progress in this direction requires the explicit development of a detailed model followed by rigorous fitting to observational data.} 


\section{Summary and Conclusions}
\textcolor{black}{The 2017 Event Horizon Telescope (EHT) observations of Sagittarius $A^*$ identified an ABH without a traditional event horizon, i.e., the JMN1 spacetime represents one of the best black hole mimickers for Sgr $A^*$ \cite{EventHorizonTelescope:2022xqj}. As stated, ``More strikingly, for the same approximate luminosity as Sgr A* at 200 GHz, accretion flows onto these singularities have spectra nearly identical to those of a Schwarzschild black hole, indicating that a JMN-1 naked singularity with a photon sphere may be one of the best possible black hole mimickers for Sgr $A^*$"\cite{EventHorizonTelescope:2022xqj}.}
Considering the various aspects of these models,
we have proposed here that the ABH scenario offers a more plausible explanation for the {sustainable} quenching of the galaxy. The two toy models, namely JMN1 \& JMN2, have been proposed as the globally naked singularities, which form through the gravitational collapse of the physically relevant matter clouds. These models are just examples among the larger group of equilibrium configurations that could arise from the gravitational collapse, throughout which the metric of the collapsing cloud assumes the forms given in Eqn. (\ref{jmn1}) \& (\ref{jmn2}). The existence of the event horizon around the astrophysical black holes has been under question since the inception of the cosmic censorship conjecture(CCC). The gravitational collapse of the matter clouds could end in various possible entities, including black holes, ultra-dense compact objects without event horizons, globally and locally naked singularities, etc. The black holes in the AGNs have always been central to explaining the quenching of the galaxy and host of other astrophysical manifestations. From Silk and Rees in 1998 to the present times, the role of BHs in galaxy quenching has been explored through various mathematical tools and simulations\cite{Cattaneo:2009ub, harrison, report}. The various shortcomings of the AGN feedback models are noted in Sect. 3, which provided us the impetus to explore the role of some alternative end-products of the gravitational collapse of the matter clouds, with apparently more realizable physical conditions.

The behavior of the innermost stable circular orbits is the key distinction between the BHs and the ABHs without EHs, and it is this attribute that makes ABHs a more effective galaxy-quenching {tool}. Because of the absence of the EHs in ABHs, the ISCOs can extend up to the center of the ultra-dense compact object, making that ultra-strong gravity region a part of the accretion disk. Thus, ABHs can produce ultra-high energy particles which can heat the interstellar medium of the host galaxy; the details of this process can be found in \cite{report} and the references therein. {We have discussed the static ABH cases and compared it with the static black hole spacetime, which is Schwarzschild black hole, for the simplicity of the argument. However the calculation can also be done for the rotating cases and can be compared with the rotating black hole, which is Kerr spacetime.}

The short-term quenching, on the other hand, can be explained through stellar-mass ABHs. When a massive star runs out of its (nuclear) fuel, it would undergo a phase of catastrophic collapse and due to the inhomogeneity of the matter in the parent cloud, the singularity, the event horizon, and the apparent horizon could form at the same comoving time\cite{Dwivedi:1992fh, Joshi:1993zg}. As they form simultaneously, {a} family of null geodesics (the light rays) exists, which could escape to infinity, making the singularity visible and observable, in principle. {During} the moments before getting covered by the event horizon, there can be a burst of energetic radiation emanating from the ultra-strong gravity region, i.e., from the vicinity of the singularity. Conceivably, {such events} could be a significant contributor to the short-term quenching of the galaxy. 

We have adopted the basic calculation of Silk and Rees to show that the ABHs will theoretically require about 17 times more mass than the BHs to expel all the material from their host galaxies. The reason lies in the $\sim 100\%$ radiative efficiency of the accretion disk around ABHs{, as explained in Section C}. 
However, we have also presented the general calculation of the radiation output in an accretion disk model and discussed that ABHs would have an increasing radiative efficiency in the {formation phase} of the ultra-dense, compact object, and it will be 100\% once the singularity is formed in the infinite comoving time. The main difference between the quenching due to the BHs and ABHs can be understood through their behaviors in the `quasar' and `radio' modes. In the `quasar' mode, when the accretion rate of an AGN is very high, the BHs are very efficient in producing the radiation-driven winds as the radiative efficiency of the BHs is itself very low (i.e., ~ 6\%), hence the requirement of a much higher accretion rate in the case of BHs.
However, an ABH is forming ultra-dense compact objects at the center, so its radiative efficiency is increasing as the ISCOs get closer and closer to the center. In fact, ABHs are efficient producers of both the radiation-driven and mechanically driven winds in the quasar mode. On the other hand, the `radio' mode differs in the two cases, as the BHs produce mechanically driven winds and/or jets with greater efficiency, whereas the ABHs are more efficient in producing radiation-driven winds. 
Still, the radiative efficiency of the ABHs is reaching the maximum.
\textcolor{black}{We have examined wind generation from the accretion disk around a Schwarzschild black hole (BH) and ABHs without an event horizon. Our analysis shows that while the escape conditions exhibit similar behavior in both cases, ABHs are more efficient at producing winds, particularly at low accretion rates, where BHs fail to do so. This efficiency enables ABHs to drive winds even with minimal accretion, contributing to the quenching of star formation.}

We have also briefly discussed the possibility of observing the {StMABHs}, which can produce energies many powers of ten above the Eddington rate, {until} they get enveloped by an event horizon. The Parker winds, having powerful electric currents, could signal this transition. This paper is a preliminary attempt to outline possible scenarios which can be tested observationally, as well as simulated for various accretion disk models, as already done for the BHs\cite{DiMatteo:2005ttp, Springel:2004kf, Croton:2005hbr, Sijacki:2007rw, Noble:2008tm, Ciotti:2009qg, Shin:2009we, Ciotti:2010tu, Gabor:2010hd, Novak:2010hc, Wagner:2012hj, Choi:2013ila, Zubovas:2014sna, Henriques:2014sga, Sijacki:2014yfa, dugan, Pontzen:2016wwf, weinbergen, Koudmani:2022ymo, Piotrowska:2021xrs}. The key difference in simulating the ABHs without EHs and the normal BHs would be in terms of radiative efficiency which is linked to the ISCO. It can therefore be expected that, for a given mass, ABHs would be more effective in galaxy quenching, compared to the normal BHs.

\textcolor{black}{
We would like to highlight several future directions motivated by this work. Active Galactic Nuclei (AGNs) are conventionally modeled with supermassive black holes. In this study, we propose an alternative possibility involving horizonless ultra-compact objects, referred to here as Astrophysical Black Holes (ABHs), which may arise from the gravitational collapse of physically relevant matter clouds\cite{Joshi:2011zm, Joshi:2013dva}. These objects could offer a different perspective on long-term quenching of star formation.
A natural next step is the explicit modeling of AGNs powered by ABHs without an event horizon. Current observations, including the Event Horizon Telescope image of Sgr~A$^*$, do not yet allow a definitive distinction between black holes and certain naked singularity configurations. Developing AGN models based on ABHs and comparing their predicted galactic effects with those of standard black hole models may therefore provide new insights into fundamental questions such as the physical existence of event horizons and the validity of the cosmic censorship conjecture.
If supported by future studies, ABH-based scenarios may also help explain long-term quenching and could imply an alternative pathway for galaxy evolution. In the context of stellar-mass ABHs\cite{Dwivedi:1992fh, Joshi:1993zg}, we have suggested potential observational signatures that may appear prior to horizon formation. Further investigation along these lines could improve our understanding of the late stages of continual gravitational collapse, a regime characterized by ultra-strong gravity where quantum effects are expected to become important.
This aspect becomes even more important in the context of recent JWST observations reporting a compact quenched galaxy at $z = 7.3$, when the universe was only about 700 million years old. Explaining such early quenching generally requires a preceding starburst event, followed by strong feedback that suppresses star formation\cite{looser}.
In this scenario, stellar-mass Astrophysical Black Holes (StMABHs) could provide the necessary feedback. Further exploration of this possibility may offer insight into quenching mechanisms in the early universe and could also shed light on the potential formation of naked singularities during epochs when the earliest galaxies became visible.
Finally, if magnetic field structures in outflows — particularly winds resembling Parker-type solutions and carrying strong electric currents — are interpreted as signatures of matter or radiation emerging from naked singularities, their detailed study may offer a phenomenological route toward probing quantum gravity effects. Such observationally guided approaches could help bridge the gap between astrophysical measurements and current theoretical models of quantum gravity.
}

{Acknowledgment: GK would like to thank the Indian National Science Academy for a Senior Scientist position during which much of this work was done.}

\end{document}